\documentclass[preprint,12pt,times,nopreprintline]{elsarticle}

\usepackage{amssymb}
\usepackage{amsmath}
\usepackage{amsthm}
\usepackage{float}
\usepackage{xcolor}
\usepackage{url}

\usepackage{tablefootnote}

\begin{document}
\emergencystretch=3em

\begin{frontmatter}



\title{Learning from all particles in high-energy collisions}

\author[inst1,inst2]{Yongfeng Zhu\fnref{equal}}

\author[inst1,inst3]{Yuexin Wang\fnref{equal}}

\author[inst4]{Hao Liang}

\author[inst1,inst5]{Yuzhi Che}

\author[inst1]{Hengyu Wang}

\author[inst1]{Jianfeng Jiang}

\author[inst2]{Chen Zhou}

\author[inst5]{Huilin Qu}

\author[inst1]{Manqi Ruan\corref{cor1}}
\ead{ruanmq@ihep.ac.cn}

\affiliation[inst1]{organization={Institute of High Energy Physics, Chinese Academy of Sciences},
            addressline={Shijingshan District}, 
            city={Beijing},
            postcode={100049}, 
            country={China}}
\affiliation[inst2]{organization={State Key Laboratory of Nuclear Physics and Technology, School of Physics, Peking University},
            addressline={Haidian District}, 
            city={Beijing},
            postcode={100871}, 
            country={China}}
\affiliation[inst3]{organization={China Spallation Neutron Source Science Center},
            city={Dongguan},
            postcode={523803}, 
            country={China}}
\affiliation[inst4]{organization={Laboratoire Leprince-Ringuet, CNRS, Ecole polytechnique, Institut Polytechnique de Paris},
            addressline={Palaiseau}, 
            city={Paris},
            postcode={91120}, 
            country={France}}            
\affiliation[inst5]{organization={China Center of Advanced Science and Technology},
            addressline={Haidian District}, 
            city={Beijing},
            postcode={100190}, 
            country={China}}

\fntext[equal]{These authors contributed equally to this work.}
\cortext[cor1]{Correspondence: ruanmq@ihep.ac.cn (M.R.)}

\begin{abstract}

Particle colliders stand as an irreplaceable pillar of inquiry for exploring the fundamental building blocks of matter and forces of the Universe, yet fully decoding complex collision event information remains a significant challenge. Recent advances in artificial intelligence (AI) have revolutionized complex data analysis across scientific disciplines, inspiring novel strategies to extract the rich information embedded in collider events. Here we introduce two complementary concepts---the holistic approach and Advanced Color Singlet Identification---to enhance signal-background separation, which is a critical prerequisite for precise physics measurements. By leveraging all reconstructed particles and inferring their parentage via deep learning, these methods improve the precision of key Higgs physics benchmark measurements by up to sixfold and enable realistic prospects for observing rare Higgs decays previously deemed inaccessible. Our results demonstrate how integrating particle-level information with modern AI technologies can substantially boost the discovery potential of high-energy colliders, paving a new path to unravel the fundamental physical laws underlying particle physics experiments.

\end{abstract}


\end{frontmatter}


\section{Introduction}

Colliders play an indispensable role in exploring the fundamental building blocks of matter and the forces of nature. 
Modern collider experiments produce enormous highly structured, high-dimensional data~\cite{ALEPH:2005ab,Morrissey:2009tf,Evans:2008zzb,Abe:2013kxa}, making physics measurements increasingly dependent on advanced data-analysis methodologies. 
Recent advances in AI have shown a remarkable ability to extract information directly from high-dimensional data~\cite{NobelPhys2024,NobelChem2024,Devlin:2018mgb,Brown:2020mpj,guo2025deepseek,jumper2021highly,abramson2024accurate,bi2023accurate,bodnar2025foundation}, and the particle physics community has begun to harness these techniques to enhance analysis performance, for example in jet flavor tagging~\cite{Qu:2019gqs,ParT} and jet origin identification~\cite{PhysRevLett.132.221802}.
Despite these advances, the information used in collider measurements remains highly compressed: analyses often rely on a limited set of expert-designed observables---typically of $\mathcal{O}(10)$---to characterize events that intrinsically contain orders of magnitude more degrees of freedom at the reconstructed-particle level. 
While effective, this compression inevitably discards a large fraction of the available information and becomes increasingly suboptimal as detector and reconstruction technologies improve~\cite{ALEPH:1994ayc,CMS:2017yfk,ILC_ILD,CEPC_CDR_Phy,CEPC_TDR_Det,Wang:2024eji}. 
This raises a question: to what extent are current analysis strategies limited not by detector performance or statistical power, but by how event information is represented?

Motivated by this observation, we explore a novel analysis concept in which collider events are represented and analyzed directly at the reconstructed-particle level, introducing the holistic approach that treats all reconstructed particles in an event as the fundamental representation for inference to discriminate the signal and background processes.
While such a holistic representation maximizes the accessible particle-level information, certain measurements remain limited by our ability to resolve the causal structure of events.
In fully hadronic final states, the assignment of reconstructed particles to their parent color-singlet systems---commonly referred to as color singlet identification (CSI)---constitutes a long-standing bottleneck for precision measurements, closely related to the QCD hadronization~\cite{Webber:1994zd}.
To address this limitation, we introduce Advanced Color Singlet Identification (ACSI), a complementary deep learning approach that infers particle-level parentage information.

Higgs boson is a central focus of nowadays particle physics, as it could be a gateway to new physics beyond the Standard Model (SM).
The electron-positron Higgs factory offers ideal experimental conditions for the Higgs measurements, and is identified as the highest priority future collider~\cite{European:2720131,deBlas:2025gyz}.
Multiple Higgs factories are therefore proposed and promoted with intensive physics studies, design efforts, and key technology R\&D~\cite{CEPC_TDR_Acc,FCC:2018evy,ILC_TDR_Sum,CLIC_CDR,Bai:2021rdg}. 
We quantify the impacts of the holistic approach and ACSI using two benchmark analyses based on the core physics measurements at the electron-positron Higgs factory: 
\begin{itemize}
    \item $H \to b\bar{b}/c\bar{c}/gg/s\bar{s}$ measurements in $\nu\bar{\nu}H$ events, showcasing the performance of the holistic approach in a clean experimental environment.
    \item $H \to b\bar{b}/c\bar{c}/gg/s\bar{s}$ measurements in $q\bar{q}H$ events, illustrating the impact and synergy between the holistic approach and ACSI in fully hadronic final states.
\end{itemize}
These benchmarks target Higgs di-jet decay modes with branching ratios summed to roughly 70\%, which are essential for determining Higgs couplings, the core parameters of the SM.
At electron-positron Higgs factories, most Higgs bosons are produced via the Higgsstrahlung process, $e^+e^- \to ZH$, with $\nu\bar{\nu}H$ and $q\bar{q}H$ channels accounting for roughly 20\% and 70\% of all $ZH$ events, respectively. 
The $\nu\bar{\nu}H$ channel is extremely clean, with nearly all reconstructed particles originating from the Higgs decay, making it ideal for studying Higgs decay properties with minimal background contamination.
A proper interpretation of $q\bar{q}H$ events is essential for the ultimate measurement precision of Higgs couplings, and this precision strongly relies on the CSI performance, in particular distinguishing Higgs decay products from those of the $Z$ boson.
Therefore, the fully hadronic $q\bar{q}H$ channel provides a simple, yet highly relevant and influential benchmark for developing innovative CSI techniques.

These benchmarks are normalized to the experimental setup of the Circular Electron-Positron Collider (CEPC)~\cite{CEPC_TDR_Acc,CEPC_TDR_Det,CEPC_CDR_Phy}, one of the proposed future Higgs factories. 
The CEPC is expected to operate at the center-of-mass energy of 240 GeV, accumulating an integrated luminosity of 21.6 ab$^{-1}$ and producing 4.1 million Higgs bosons over one decade. 
All samples used in this study are generated with fast-simulation tools~\cite{deFavereau:2013fsa} that model the essential detector and reconstruction features.

\section{Materials and methods}
\label{matmet}
\renewcommand{\dblfloatpagefraction}{.9}
\begin{figure*}[t]
    \centering
    \includegraphics[width=.8\textwidth]{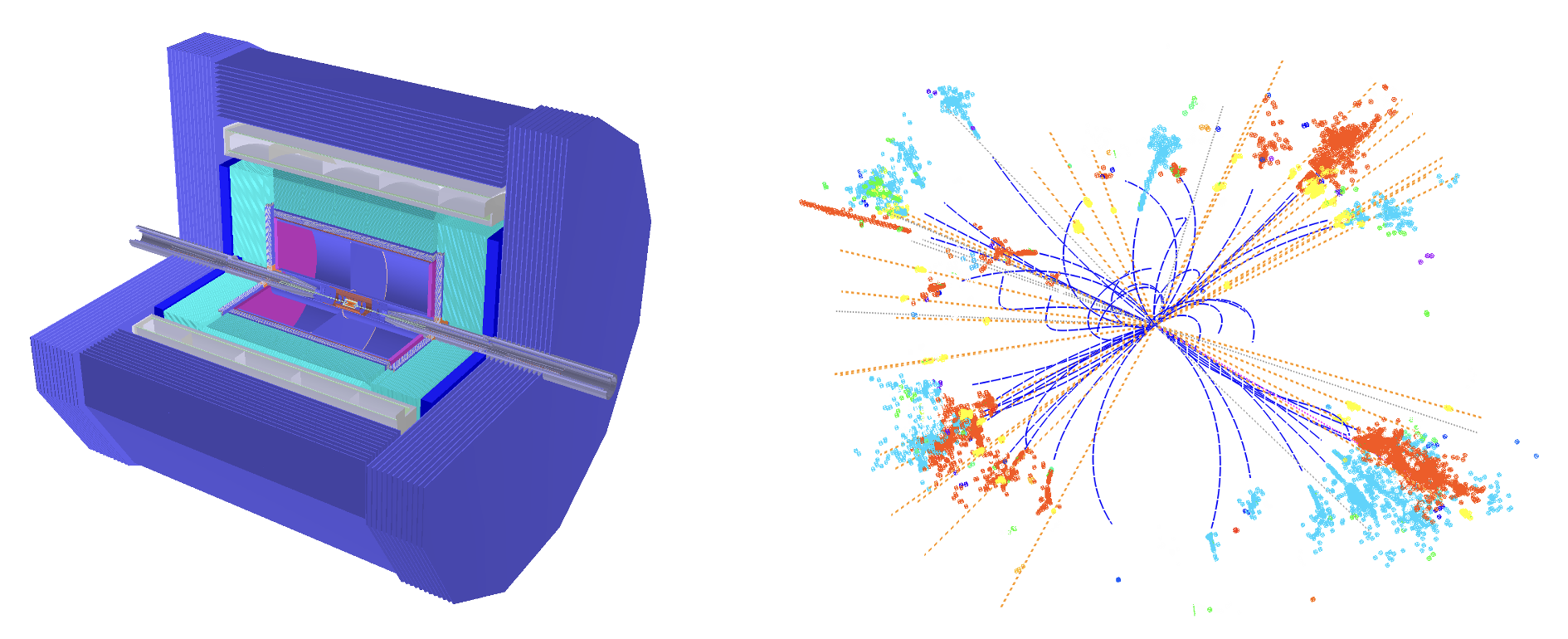}
    \caption{Geometry of the AURORA detector (left) and an event display of a reconstructed $e^+e^- \to Z(\to q\bar{q})H(\to b\bar{b})$ event at the center-of-mass energy of 240\,GeV (right). Different particles are depicted with colored curves and straight lines: \textcolor{red}{red} for $e^{\pm}$, \textcolor{cyan}{cyan} for $\mu^{\pm}$, \textcolor{blue}{blue} for $\pi^{\pm}$, \textcolor{orange}{orange} for photons, and \textcolor{magenta}{magenta} for neutral hadrons.}
    \label{fig:display}
\end{figure*}

\subsection{Detector}
The AURORA detector concept~\cite{Wang:2024eji}, evolved from the CEPC Conceptual Design Report (CDR) baseline~\cite{CEPC_CDR_Phy}, is used as the reference detector in this study. 
It features high-granularity electromagnetic and hadronic calorimeters, a low-material high-precision tracker, a high-resolution vertex detector, and a large solenoid magnet.
Fig.~\ref{fig:display} provides an overview of the AURORA detector geometry together with a representative reconstructed event.
A key design of AURORA is the usage of 5-dimensional (5D) calorimetry, where each calorimeter cell provides time measurement with a resolution of roughly 100~ps. 
Using AI-enhanced reconstruction, the AURORA concept provides a proof of principle for one-to-one correspondence reconstruction~\cite{Wang:2024eji}, which pursues confusion-free mapping between reconstructed and true final-state particles. This concept enables efficient and accurate identification of all final-state particles in a jet event.
Through the effective identification and removal of fake particles (a major confusion effect), the relative mass resolution (BMR) for hadronically decaying bosons is improved to 2.7\%~\cite{Wang:2024eji}, thereby substantially enhancing the separation among $H$, $W$, and $Z$ bosons, which is crucial for precise Higgs measurements.
In this study, we model the AURORA performance using a fast simulation tool that has been validated by the full simulation performance studies.

\subsection{Samples}
The event samples used in this study are generated using Whizard-1.95~\cite{Whizard} and Pythia-6.4~\cite{Pythia6}. 
To estimate the uncertainties introduced by hadronization, we also compare to Herwig-7.2.2~\cite{Bahr:2008pv,Bellm:2015jjp} and Pythia-8.313~\cite{Bierlich:2022pfr}.

For both $\nu\bar{\nu}H$ and $q\bar{q}H$ benchmarks, six event categories are designed. 
The former includes $\nu\bar{\nu}H(b\bar{b})$, $\nu\bar{\nu}H(c\bar{c})$, $\nu\bar{\nu}H(s\bar{s})$, $\nu\bar{\nu}H(u\bar{u})$, $\nu\bar{\nu}H(d\bar{d})$, and $\nu\bar{\nu}H(gg)$ modes, and the latter includes $WW$, $ZZ$, $q\bar{q}H(b\bar{b})$, $q\bar{q}H(c\bar{c})$, $q\bar{q}H(s\bar{s})$, and $q\bar{q}H(gg)$ modes. 
For each kind of physics process, one million events are generated to train these deep-learning models. 
An independent testing data set is generated, with statistics of roughly half a million for each physics process. 
The trained models are applied to the testing data set to extract the results by normalizing the event statistic to 
the SM prediction and the nominal integrated luminosity. 

\begin{figure*}[t]
    \centering
     \includegraphics[width=.9\textwidth]{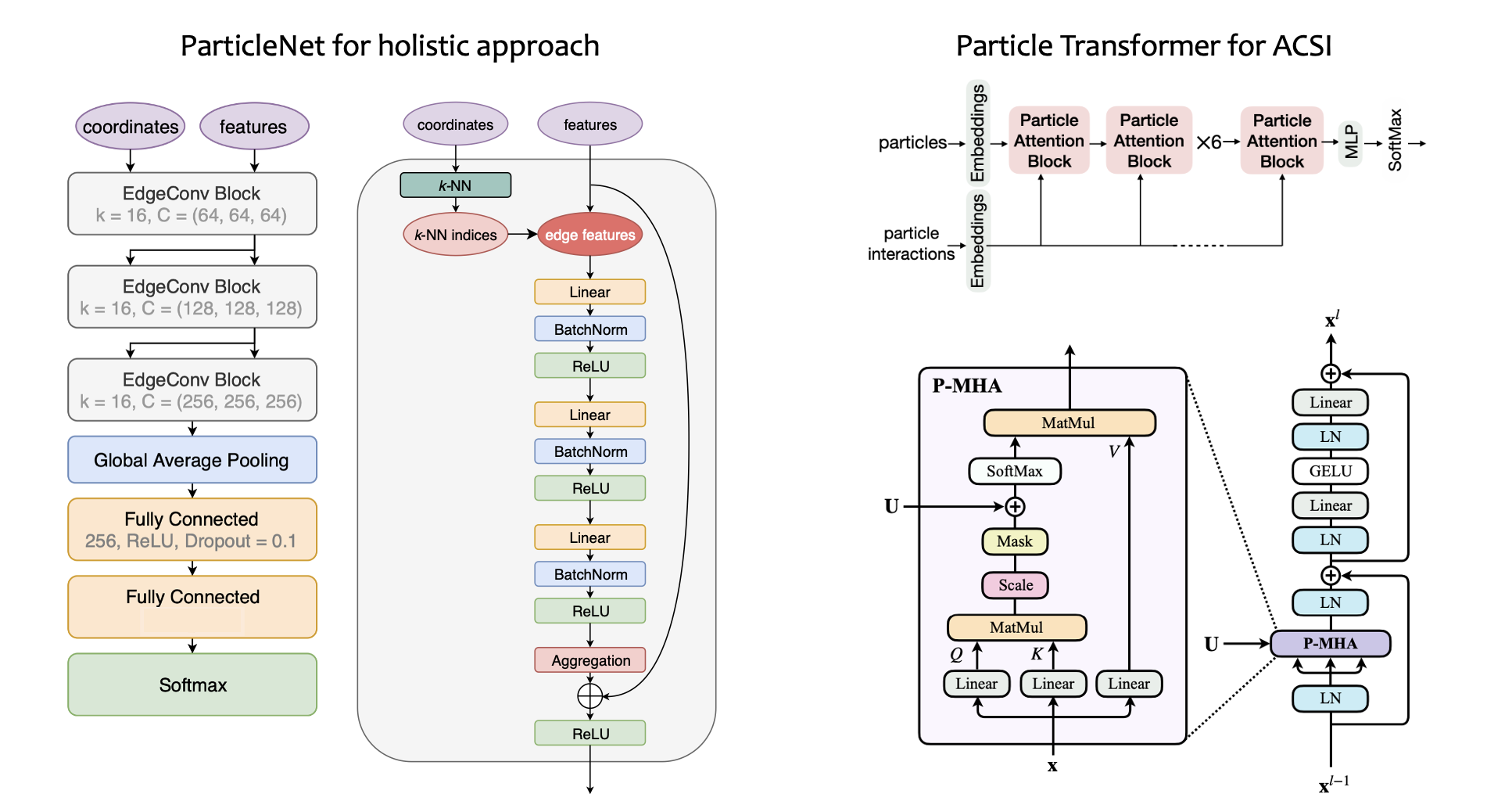}
    \caption{\label{fig:model} \textbf{Model architectures.} Left: ParticleNet architecture and the EdgeConv block for the holistic approach.
    Right: Particle Transformer architecture for ACSI composed of eight particle–attention blocks followed by a multilayer perceptron for particle-level classification.}
\end{figure*}

\subsection{Methodology of holistic approach}
For the holistic approach, we employ ParticleNet~\cite{Qu:2019gqs} for event classification. 
The overall structure of ParticleNet is illustrated in Fig.~\ref{fig:model}. 
It consists of three EdgeConv blocks, followed by a channel-wise global average pooling layer and two fully connected layers, culminating in a softmax output that provides event-level classification probabilities.
The key innovation of ParticleNet lies in the EdgeConv operation, which performs feature aggregation over each particle and its k-nearest neighbors in the learned feature space. The workflow of a typical EdgeConv block is shown in Fig.~\ref{fig:model}. 
For each particle, the k-nearest neighbors are first identified using either spatial coordinates or learned feature embeddings, depending on the stage of the network. 
Edges are then constructed between the particle and its neighbors, and their associated features are transformed through a series of multilayer perceptrons. 
The particle features are subsequently updated by aggregating the transformed edge features, enabling the network to learn hierarchical, geometry-aware representations of the event.
The full model contains approximately 360k trainable parameters and is trained for 30 epochs in this analysis.
The version that achieves the highest validation precision during training is selected to characterize the performance.

\subsection{Methodology of ACSI}

\begin{figure*}[t]
    \centering
    \includegraphics[width=0.85\textwidth]{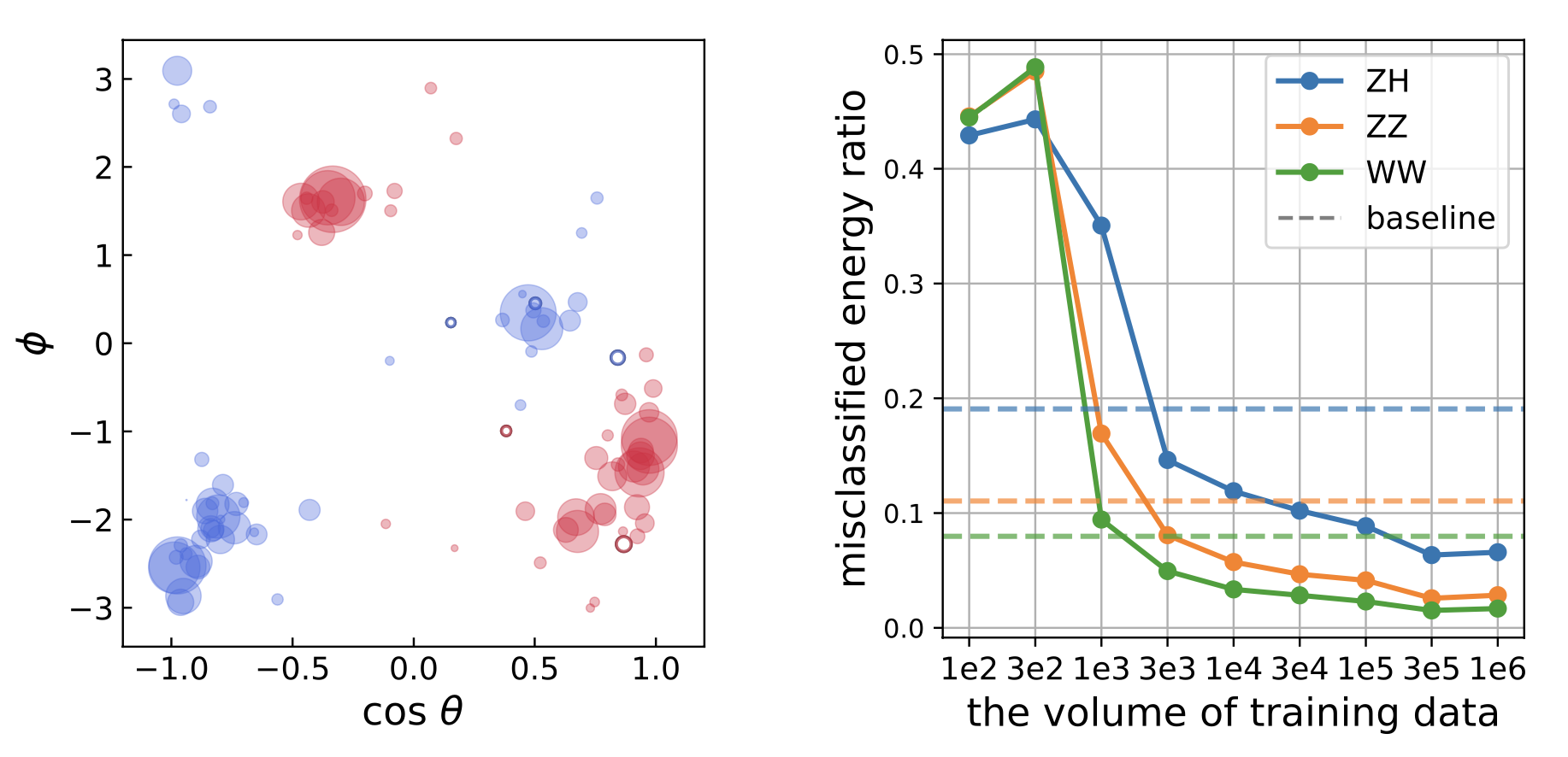}
    \caption{\label{fig:event}\textbf{ACSI performance.} 
    Left: Visualization of particle-level parentage inference with ACSI for a fully hadronic $ZH$ event at 240 GeV. Each circle represents a reconstructed final-state particle, with the area proportional to its energy. Colors indicate the parent assignment inferred by ACSI. Particles whose parent is misidentified are highlighted using open-circle markers. Right: Scaling behavior of ACSI performance as a function of the training data volume.}
\end{figure*}

The ACSI is trained on and applied to fully hadronic $WW$, $ZZ$, and $ZH$ events. 
The goal is to partition the final-state particles into two groups, each corresponding to a color-singlet boson. 
After training, ACSI assigns to each reconstructed particle two float numbers, which sum to unity and represent the inferred probabilities of originating from either boson. 
These probabilities are subsequently incorporated as particle-level features in the holistic approach, thereby enhancing the discrimination performance among different physics processes.
Conceptually, ACSI can be viewed as a ``coloring game'', in which reconstructed particles are assigned different ``colors'' according to their parent color-singlet systems, as illustrated in the left panel of Fig.~\ref{fig:event}.

The ACSI is implemented using the Particle Transformer~\cite{ParT} with approximately $2\times 10^6$ trainable parameters, which is originally designed for jet tagging. 
The Particle Transformer comprises eight particle attention blocks, each equipped with eight attention heads, followed by two additional attention layers that aggregate per-particle embeddings into a single jet-level representation for classification.
In contrast, the ACSI task requires predicting the parent boson of each final-state particle. 
Therefore, the global aggregation stage is omitted. 
Instead, we retain the per-particle embeddings after the eight attention blocks and apply a sequence of multilayer perceptrons to perform particle-level classification directly. 
The architecture of ACSI is shown in the right panel of Fig.~\ref{fig:model} with a total of approximately two million trainable parameters.
The details about the particle attention block and particle interactions could be found in the reference~\cite{ParT}.

\renewcommand{\dblfloatpagefraction}{.9}

The performance of ACSI can be characterized by the misclassified energy ratio, defined as the fraction of event energy carried by particles incorrectly assigned to their parent color-singlet systems.
When trained on samples of $WW$, $ZZ$, and $ZH$ events with equal statistics of one million each, ACSI achieves misclassified energy ratios of 1.7\%, 2.8\%, and 6.6\%, respectively, improved by 3--4 times from that of 7.9\%, 11\%, and 19\% using the conventional method ~\cite{Zhu:2022lzv}.
Furthermore, ACSI exhibits an obvious scaling behavior, where performance strongly depends on the training data size, as illustrated in the right panel of Fig.~\ref{fig:event}.
A distinct performance boost is observed at a training data size of $10^3$, a phenomenon akin to a ``phase transition'' in AI training, highlighting the importance of data-volume scaling in particle-level AI analyses.

\section{Results}

\subsection{$H \to b\bar{b}/c\bar{c}/gg/s\bar{s}$ measurements in $\nu\bar{\nu}H$ events}

This benchmark targets the discrimination of different hadronic Higgs decay modes and the precise determination of their branching ratios. 
The analysis is based on $\nu\bar{\nu}H$ event samples with $H \to b\bar{b}$, $c\bar{c}$, $gg$, and $s\bar{s}$ decays, corresponding to expected yields of $3.7\times10^{5}$, $1.8\times10^{4}$, $5.2\times10^{4}$, and $1.5\times10^2$ events, normalized to the SM branching fractions and the CEPC nominal integrated luminosity~\cite{CEPC_TDR_Acc}.
In this benchmark, SM backgrounds are neglected, as the $\nu\bar{\nu}H$ channel at electron–positron Higgs factories is extremely clean---conventional reconstruction and analysis techniques can already achieve a total $\nu\bar{\nu}H$ signal efficiency and purity of approximately 34\% and 73\%~\cite{Zhu:2022lzv}, and future advances in detector design and analysis methodologies are expected to further enhance the signal–background separation.

\begin{figure*}[!htbp]
    \centering
    \hspace*{0.05cm}
    \includegraphics[width=0.8\textwidth]{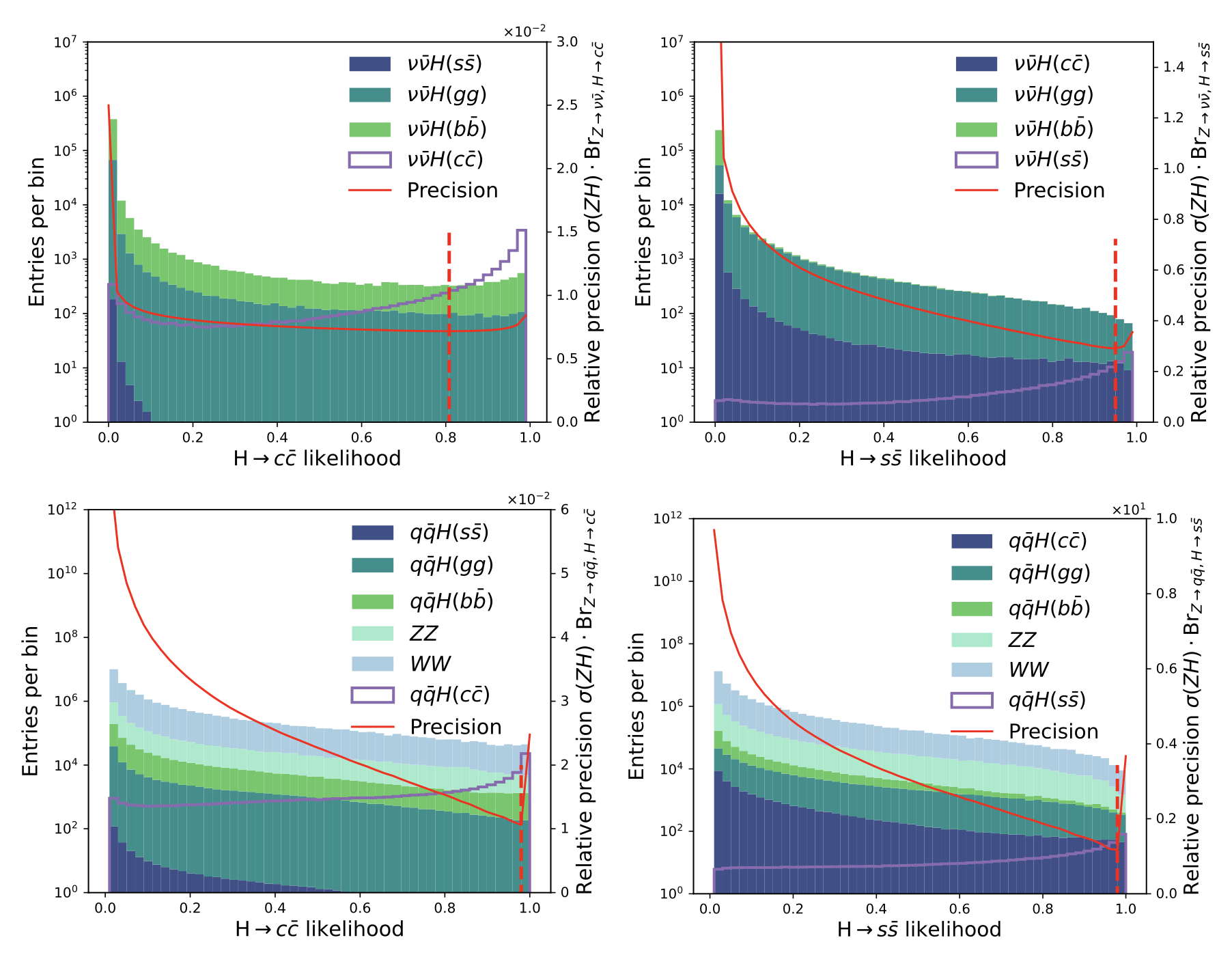}
    \caption{\textbf{Measurement performance of $H \to c\bar{c}$ and $H \to s\bar{s}$ decays with the holistic approach and ACSI.} Signal score distributions for $H \to c\bar{c}$ and $H \to s\bar{s}$ in $\nu\bar{\nu} H$ (upper panels) and $q\bar{q} H$ (lower panels) events. For the $\nu\bar{\nu} H$ channel, the optimized cut (vertical dashed line) yields signal strength precisions of 0.72\% and 29\% for $H \to c\bar{c}$ and $H \to s\bar{s}$, respectively. For the $q\bar{q} H$ channel, using the holistic approach combined with ACSI, the corresponding precisions are 1.03\% and 114\%.}
    \label{fig:sig_score}
\end{figure*}

The holistic approach takes as input reconstructed particle information, including the four-momentum, particle type, track impact parameters, and infers the relevant scores for different physics events. 
It is realized using a ParticleNet-based model~\cite{Qu:2019gqs} with approximately $3.6\times10^{5}$ trainable parameters.
Details of the model architecture and training procedure are provided in Section~\ref{matmet}.
As illustrated in Fig.~\ref{fig:sig_score}, optimized score-based selections—determined to maximize statistical precision, quantified by the relative precision of signal strength $\sqrt{N_S + N_B}/N_S$, where $N_S$ and $N_B$ represent the number of signal and background events—yield relative precisions of 0.14\%, 0.72\%, 0.46\%, and 29\% for $H \to b\bar{b}$, $c\bar{c}$, $gg$, and $s\bar{s}$ channels.
Compared with existing analyses based on cut-based selections and Boosted Decision Trees (BDT)~\cite{friedman2001greedy}, the holistic approach achieves a two- to six-fold improvement in precision at the same luminosity~\cite{Zhu:2022lzv}. 
This gain reflects the ability of the holistic approach to exploit the full breadth of information at the reconstructed-particle level, which is largely omitted in conventional analyses.

\renewcommand{\dblfloatpagefraction}{.9}


\subsection{$H \to b\bar{b}/c\bar{c}/gg/s\bar{s}$ measurements in $q\bar{q}H$ and fully hadronic $ZZ$/$WW$ events}

This benchmark is designed to separate six fully hadronic physics processes, including the Higgs signal processes $q\bar{q}H$ with $H \to b\bar{b}/c\bar{c}/gg/s\bar{s}$ decays and the fully hadronic $ZZ$ and $WW$ background processes.
The expected event yields are $1.1\times10^6$, $5.4\times10^5$, $1.5\times10^5$, $4.5\times10^2$, $1.1\times10^7$, and $1.67\times10^8$, respectively, according to the CEPC nominal configuration~\cite{CEPC_TDR_Acc}.
Other fully hadronic background processes are neglected for simplicity, as they are either subdominant or can be efficiently identified~\cite{Zhu:2022lzv}.

CSI is essential for precision measurements in this benchmark, and more generally for physics measurements involving fully hadronic final states.
Compared to $\nu\bar{\nu}H$ events, $q\bar{q}H$ events contain nearly twice as many final-state particles, resulting in a substantially more complex event topology.
Meanwhile, the statistics of $ZZ$ and $WW$ backgrounds are 1--2 orders of magnitude larger than the total Higgs signals.
Detectors at electron–positron Higgs factories are expected to deliver a relative mass resolution of approximately 2.7--4\% for hadronically decaying $Z$, $W$, and Higgs bosons~\cite{An:2018dwb,Wang:2024eji}; therefore, if the origin of each final-state particle could be correctly identified, $ZZ$ and $WW$ events could be efficiently separated from the $ZH$ signal.
In practice, however, any mis-grouping of final-state particles can easily spoil this separation and degrade the signal purity.
In the conventional realization of CSI using jet clustering and matching, the resulting ambiguities become a major bottleneck that severely limits the precision of physics measurements in fully hadronic events~\cite{Zhu:2022lzv, CEPC_Higgs_snowmass2022}.

\renewcommand{\dblfloatpagefraction}{.9}
\begin{figure*}[t]
    \centering
     \includegraphics[width=.98\textwidth]{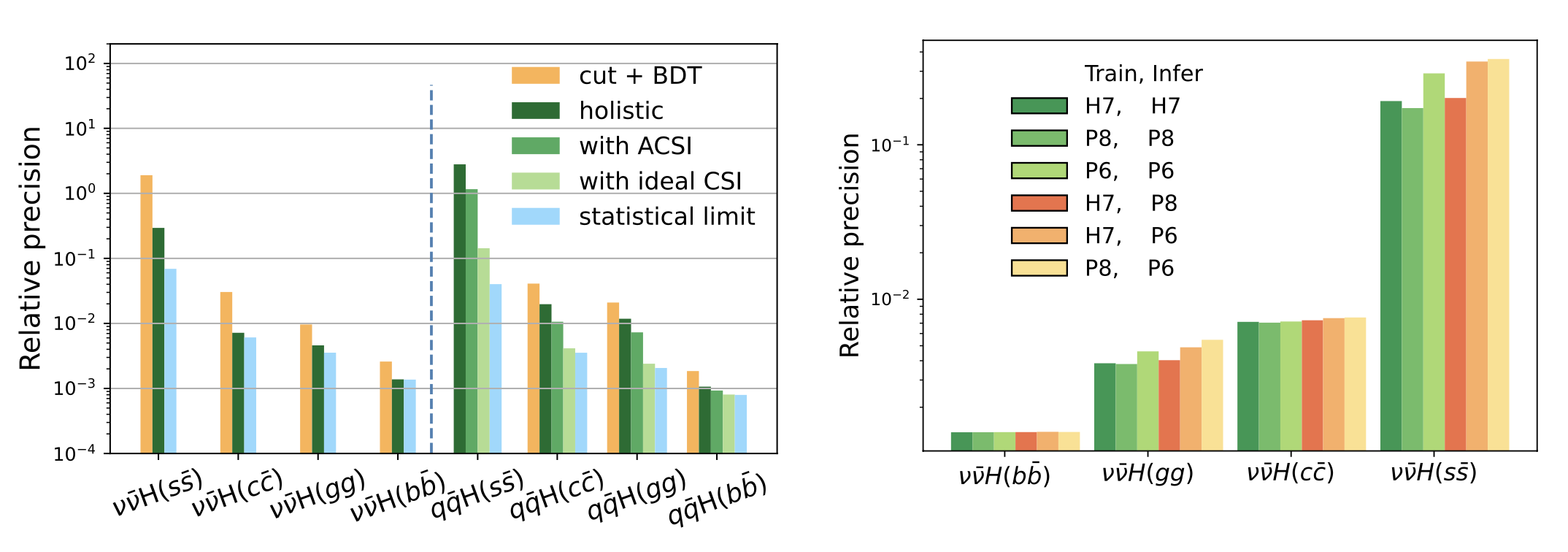}
    \caption{\label{fig:summary} \textbf{Left: Comprehensive summary of signal strength measurements.} A comprehensive summary of signal strength measurements of $q\bar{q}H(j\bar{j})$ and $H \to b\bar{b}/c\bar{c}/gg/s\bar{s}$ in $\nu\bar{\nu}H$ and $q\bar{q}H$ channels. Results for the $H \to b\bar{b}$, $c\bar{c}$, and $gg$ using cut-based and BDT methods, which include a more comprehensive set of backgrounds, are taken from Ref.~\cite{Zhu:2022lzv}, while the result for $H \to s\bar{s}$ with cut-based and BDT is taken from Ref.~\cite{PhysRevLett.132.221802}.
    \textbf{Right: Theoretical uncertainty from hadronization models.} The $H \to b\bar{b}/c\bar{c}/gg/s\bar{s}$ measurement in $\nu\bar{\nu}H$ channel with different hadronization models, where P6, P8, and H7 represent Pythia-6.4~\cite{Pythia6}, Pythia-8.313~\cite{Bierlich:2022pfr}, and Herwig-7.2.2~\cite{Bahr:2008pv,Bellm:2015jjp}, respectively.}
\end{figure*}

\renewcommand{\dblfloatpagefraction}{.9}
\begin{table*}[t]
    \centering
    {
    \scriptsize
    \setlength{\tabcolsep}{2.5pt}
    \renewcommand{\arraystretch}{1.2}
    \caption{\textbf{Demonstration of the effectiveness of the holistic approach and ACSI through signal strength precision.} When the CEPC operates as a Higgs factory and collects an integrated luminosity of 20 ab$^{-1}$, the signal strength precisions for the decay modes $H \to b\bar{b}/c\bar{c}/gg/s\bar{s}$ in $\nu\bar{\nu}H$ and $q\bar{q}H$ channels are evaluated under five different scenarios. The bottom four rows present results assuming only irreducible backgrounds---specifically, Higgs decays themselves in the $\nu\bar{\nu}H$ channel, and fully hadronic $WW$ and $ZZ$ events in the $q\bar{q}H$ channel. For comparison, the conventional approach refers to cut-flow followed by BDT for event classification.}    
    \label{tab:Hccggacc}    
    \setlength{\tabcolsep}{3.pt}

    \resizebox{\textwidth}{!}{%
    \begin{tabular}{c|cccc|cccc}
    \hline
    & \multicolumn{4}{c|}{$\nu\bar{\nu}H$}  
    & \multicolumn{4}{|c}{$q\bar{q}H$}  \\
    
    & $H \to b\bar{b}$ & $H \to c\bar{c}$ & $H \to gg$ & $H \to s\bar{s}$ 
    & $H \to b\bar{b}$ & $H \to c\bar{c}$ & $H \to gg$ & $H \to s\bar{s}$ \\
    \hline
    cut + BDT   
    & 0.26\%\cite{Zhu:2022lzv} & 3.04\%\cite{Zhu:2022lzv} & 0.96\%\cite{Zhu:2022lzv} & 190.00\%\cite{PhysRevLett.132.221802}
    & 0.19\%\cite{Zhu:2022lzv}  & 4.10\%\cite{Zhu:2022lzv}  & 2.10\%\cite{Zhu:2022lzv} & -\\
    
    holistic
    & 0.14\% & 0.72\% & 0.46\% & 29.34\% 
    & 0.11\% & 1.96\% & 1.05\% & 279\% \\
    
    holistic with CSI
    &- &- &- &- 
    & 0.09\% & 1.03\% & 0.73\% & 114\% \\
    
    holistic with ideal CSI
    &- &- &- &- 
    & 0.08\% & 0.41\% & 0.24\% & 14.32\% \\
    
    statistical limit       
    & 0.14\% & 0.61\% & 0.36\% & 6.91\% 
    & 0.08\% & 0.35\% & 0.21\% &4.02\% \\

    \hline
    improvement ratio
    & 1.9   & 4.2     & 2.1     & 6.5 
    & 2.1     & 4.0     & 2.9     & - \\
    \hline
    \end{tabular}
    }
    \par\vspace{0.5ex}\raggedright\scriptsize
    Note: For $\nu\bar{\nu}H$ events, the improvement ratio is defined as
    $\frac{\text{cut}+\text{BDT}}{\text{holistic}}$, while for $q\bar{q}H$
    events it is defined as $\frac{\text{cut}+\text{BDT}}{\text{holistic with CSI}}$.
    }
\end{table*}

To address this limitation, we introduce ACSI, a deep-learning-based approach that infers particle-level parentage information.
We compare the anticipated precisions of four Higgs measurements with different analysis methodologies, as summarized in Table~\ref{tab:Hccggacc} and Fig.~\ref{fig:summary}.
Using BDT, the conventional methodology predicts that $H\to b\bar{b}, c\bar{c}, gg, s\bar{s}$ could be measured to relative precisions of 0.26\%, 3.04\%, 0.96\%, 190\%, respectively, scaled from a dedicated full-simulation study~\cite{Zhu:2022lzv}.
The holistic approach, by leveraging the high dimensionality of reconstructed-particle-level information, improves the anticipated precisions of $H\to b\bar{b}, c\bar{c}, gg$ measurements by roughly a factor of two.
For each reconstructed particle, ACSI provides two additional scores, representing the probabilities that the particle originates from one of the bosons or the other.
Incorporating this information into the holistic approach, the anticipated precisions are further improved by 20\% for $H \to b\bar{b}, gg$, and by factors of 2--3 for $H \to c\bar{c}, s\bar{s}$.

For reference, we also consider an idealized scenario in which final-state particles are perfectly assigned to their parent color-singlet systems.
In this limit, the holistic analysis approaches the statistical precision bound for the $H \to b\bar{b}$, $c\bar{c}$, and $gg$ measurements, reaching within approximately 10\% of the statistical limit, while the $H \to s\bar{s}$ channel remains far from the statistical limit, primarily due to its small branching ratio and thus much more vulnerable to background contamination.
The performance achieved with ACSI lies approximately halfway between the holistic-only and the ideal CSI scenarios, 
demonstrating its effectiveness in resolving the critical parentage information. 
Meanwhile, further exploration toward the statistical limit becomes an intriguing and challenging topic.

\section{Discussion}

It should be noted that the benchmark analyses presented in this manuscript adopt a simplified approach. 
In real experiments, the final measurement precision will be influenced by reducible backgrounds, systematic uncertainties arising from detector imperfections and instabilities, and theoretical uncertainties stemming from discrepancies between real-world physics and the relevant theoretical models, etc.
Explicitly, we would like to point out:

\begin{itemize}
    \item \emph{Backgrounds:} The current analysis considers only irreducible backgrounds. 
    However, reducible backgrounds—such as two-fermion processes, di-photon events, beam-induced backgrounds, and even detector noise—are also relevant to those measurements. 
    At a Higgs factory like CEPC, key measurements like $\ell^+\ell^-H$ (Higgs recoil) and $\nu\bar{\nu}H$ with $H \to b\bar{b}/c\bar{c}/gg$ typically achieve signal efficiencies times purities ranging from 25\% to 80\%~\cite{chen2017cross,Zhu:2022lzv,Bai_2020}, even with conventional cut-based and BDT approaches. 
    In addition, the holistic and ACSI proposed in this study are expected to further enhance the separation between signal and reducible background.
    Therefore, we expect the impact of reducible backgrounds to be moderate and controllable.
    \item \emph{Experimental systematic uncertainties:} Experimental systematic uncertainties arise from limited understanding of the detector, with imperfections and instabilities in the detector system potentially playing a critical—and even dominant—role. 
    Systematic control is thus a key requirement in detector research and design for future Higgs factories.  
    In this context, the development of detectors capable of one-to-one correspondence between reconstructed particles and final-state particles, as proposed in Ref.~\cite{Wang:2024eji}, could be essential to control the relative systematics, by providing the inclusive information of reconstructed particles in the continuous data stream.
    \item \emph{Theoretical uncertainties:} Both the holistic and ACSI approaches are based on supervised learning, which relies on high-quality and well-validated simulations. 
    Since these methods use inclusively reconstructed particles as input, the modeling of hadronization~\cite{Webber:1994zd}, which governs the distribution of final-state particles, becomes critical. 
    To assess the impact of hadronization models, we extract results under several scenarios. 
    Each scenario involves a pair of hadronization models—one for training and one for inference. 
    The results are summarized in Fig.~\ref{fig:summary}.
    For the $H \to b\bar{b}$ and $H \to c\bar{c}$ channels, different combinations yield comparable performance, demonstrating the robustness of the holistic approach.
    For $H \to gg$ and $H \to s\bar{s}$, variations between model combinations can result in relative differences of up to a factor of two—likely due to the less well-understood hadronization processes of gluon and strange jets in current models.
    This discrepancy underscores the need for more accurate and data-driven hadronization modeling, which can be improved iteratively through validation with real experimental data.
    Combinations involving Herwig-7.2.2 and Pythia-8.313 consistently outperform those with Pythia-6.4, as the former two predict higher yields of charged kaons, enhancing the identification of strange jets.
    It is also notable that the results with the same model in training and inference always lead to slightly better results compared to the case of changing one model to an alternative.
    \item  \emph{$H\to s\bar{s}$ measurement:} Another key message of this study is the potential to observe the rare decay $H \to s\bar{s}$ at future Higgs factories.
    In the $\nu\bar{\nu}H$ channel with only $H \to j\bar{j}$ backgrounds considered, our analysis demonstrates that the $H \to s\bar{s}$ signal could be measured with a relative precision of 17\% using Pythia-8313, 19\% using Herwig-7.2.2, and 29\% using Pythia-6.4 hadronization models\footnote{Herwig-7.2.2 and Pythia-8.313 generate more $K^{\pm}$ mesons in the final state than Pythia-6.4, which enhances discrimination of the strange-quark signal.}.
    \end{itemize}

\section{Conclusion}

In summary, we propose and realize two complementary concepts---the holistic approach and ACSI---that substantially enhance event classification in experimental particle physics through deep learning.
By exploiting inclusive reconstructed-particle-level information, these concepts increase the effective dimensionality of the input by roughly two orders of magnitude compared to conventional analyses, resulting in two- to six-fold improvements in the anticipated statistical precision of Higgs measurement benchmarks.
In fully hadronic final states, ACSI mitigates the long-standing limitation of conventional jet-based CSI by resolving particle-level parentage, providing an additional improvement of approximately a factor of two.
Together, these concepts yield an overall enhancement in discovery power of roughly a factor of three, corresponding to an effective increase in integrated luminosity by about one order of magnitude.
A particularly notable outcome is the markedly improved sensitivity to the rare decay $H\to s\bar{s}$, 
rendering this channel accessible at the level of $\mathcal{O}(20\%)$ precision in the $\nu\bar{\nu}H$ channel under different hadronization models, with further gains expected when additional Higgs production channels are included.
While this study relies on supervised learning and accurate simulations, continued improvements in simulation tuning and data-driven validation will be essential for controlling systematic uncertainties, and the proposed concepts are general and applicable to a broad class of collider processes characterized by reconstructed final-state particles.

\section*{Declaration of interests}

The authors declare no competing interests.

\section*{Ethical statement and patient consent}
This study did not involve human participants, human-derived materials, patient data, or animal experiments. Ethical approval and patient consent were therefore not required.

\section*{Data and code availability}
The datasets and code used in this study are not publicly available. They are available from the corresponding author upon reasonable request.

\section*{Supplemental information}
No supplemental information is provided with this manuscript.

\section*{Declaration of AI and AI-assisted technologies in the writing process}
During manuscript preparation, AI-assisted tools were used to assist with language polishing and formatting. The authors reviewed and edited the manuscript and take full responsibility for its content.

\section*{Funding and acknowledgments}
This work was partially supported by the China Postdoctoral Science Foundation under Grant Number 2025M773403 and the National Natural Science Foundation of China under grant No. 12342502. The funders had no role in study design, data collection and analysis, decision to publish, or preparation of the manuscript.

\section*{Author contributions}
M.R. initiated, designed, and supervised the project. H.Q. designed the artificial intelligence models. Y.Z. conducted the numerical experiments. H.Q., Y.C., H.W., J.J., and C.Z. processed the data. Y.W. integrated the results. All authors prepared the manuscript and discussed the research.

\section*{Lead contact}
Further information and requests should be directed to Manqi Ruan (\url{ruanmq@ihep.ac.cn}).


\begin{thebibliography}{00}

\bibitem{ALEPH:2005ab}
ALEPH, DELPHI, L3, OPAL, SLD, LEP Electroweak Working Group, SLD Electroweak Group, SLD Heavy Flavour Group.
Precision electroweak measurements on the Z resonance.
Phys. Rept. 2006;427:257--45.

\bibitem{Morrissey:2009tf}
Morrissey DE, Plehn T, Tait TMP.
Physics searches at the LHC.
Phys. Rept. 2012;515:1--113.
DOI: 10.1016/j.physrep.2012.02.007.

\bibitem{Evans:2008zzb}
Evans L, Bryant P.
LHC Machine.
J. Instrum. 2008;3:S08001

\bibitem{Abe:2013kxa}
Abe T, et al.
Achievements of KEKB.
PTEP 2013;2013:03A001.

\bibitem{NobelPhys2024}
The Nobel Prize in Physics 2024.
\url{https://www.nobelprize.org/prizes/physics/2024/summary}.

\bibitem{NobelChem2024}
The Nobel Prize in Chemistry 2024.
\url{https://www.nobelprize.org/prizes/chemistry/2024/summary}.

\bibitem{Devlin:2018mgb}
Devlin J, Chang MW, Lee K, et al.
BERT: Pre-training of Deep Bidirectional Transformers for Language Understanding.
In: Proceedings of the 2019 Conference of the North American Chapter of the Association for Computational Linguistics: Human Language Technologies, Volume 1.
Association for Computational Linguistics; 2019. p. 4171--4186.
DOI: 10.18653/v1/N19-1423.

\bibitem{Brown:2020mpj}
Brown TB, et al.
Language Models are Few-Shot Learners.
Adv. Neur. Inf. Proc. Syst. 2020;33:1901

\bibitem{guo2025deepseek}
Guo DY, Yang DJ, Zhang HW, et al.
Deepseek-r1 incentivizes reasoning in llms through reinforcement learning.
Nature 2025;645:633--638

\bibitem{jumper2021highly}
Jumper J, Evans R, Pritzel A, et al.
Highly accurate protein structure prediction with AlphaFold.
Nature 2021;596:583--589.
 
\bibitem{abramson2024accurate}
Abramson J, Adler J, Dunger J.
Accurate structure prediction of biomolecular interactions with AlphaFold 3.
Nature 2024;630:493--500.

\bibitem{bi2023accurate}
Accurate medium-range global weather forecasting with 3D neural networks.
Bi KF, Xie LX, Zhang HH, et al.
Nature 2023;619:533--538.


\bibitem{bodnar2025foundation}
Bodnar C, Bruinsma WP, Lucic A, et al.
A foundation model for the Earth system.
Nature 2025;641:1180--1187.

\bibitem{Qu:2019gqs}
Qu, Huilin and Gouskos, Loukas.
ParticleNet: Jet Tagging via Particle Clouds.
Phys. Rev. D. 2020;101:056019

\bibitem{ParT}
Qu HL, Li CQ, Qian ST, et al.
Particle Transformer for Jet Tagging.
Proc. Mach. Learn. Res. 2022;162:18281--18292.

\bibitem{PhysRevLett.132.221802}
Liang H, Zhu YF, Wang YX, et al. 
Jet-Origin Identification and Its Application at an Electron-Positron Higgs Factory.
Phys. Rev. Lett. 2024;132:221802.
 
\bibitem{ALEPH:1994ayc}
ALEPH.
Performance of the ALEPH detector at LEP.
Nucl. Instrum. Meth. A. 1995;360:481--506.

\bibitem{CMS:2017yfk}
CMS.
Particle-flow reconstruction and global event description with the CMS detector.
J. Instrum. 2017;12:P10003.

\bibitem{ILC_ILD}
Linear Collider ILD Concept Group.
The International Large Detector: Letter of Intent.
FERMILAB-PUB-09-682-E; 2010.
DOI: 10.2172/975166.

\bibitem{CEPC_CDR_Phy}
The CEPC Study Group.
CEPC Conceptual Design Report: Volume 2 - Physics and Detector.
IHEP-CEPC-DR-2018-02, IHEP-EP-2018-01, IHEP-TH-2018-01; 2018.

\bibitem{CEPC_TDR_Det}
CEPC Study Group.
CEPC Technical Design Report - Reference Detector.
IHEP-CEPC-DR-2025-01, IHEP-EP-2025-01; 2025.

\bibitem{Wang:2024eji}
Wang YX, Liang H, Zhu YF, et al.
One-to-one correspondence reconstruction at the electron-positron Higgs factory.
Computer Physics Communications 2025;0010-4655.


\bibitem{Webber:1994zd}
Webber BR.
Summer School on Hadronic Aspects of Collider Physics.
CAVENDISH-HEP-94-17.

\bibitem{European:2720131}
The European Strategy Group.
Deliberation document on the 2020 Update of the European Strategy for Particle Physics.
CERN-ESU-014.

\bibitem{deBlas:2025gyz}
de Blas J, Dunford M, Bagnaschi E, et al. 
Physics Briefing Book: Input for the 2026 update of the European Strategy for Particle Physics.
CERN Yellow Rep. Monogr. 2025;8.
DOI: 10.23731/CYRM-2025-008.

\bibitem{CEPC_TDR_Acc}
CEPC Study Group.
CEPC Technical Design Report: Accelerator.
Radiat. Detect. Technol. Methods 2024;8:1-1105.
  
\bibitem{FCC:2018evy}
FCC.
Future Circular Collider Conceptual Design Report Volume 2.
Eur. Phys. J. ST 2019;228:261--623.

\bibitem{ILC_TDR_Sum}
Behnke T, Brau JE, Foster B, et al. 
The International Linear Collider Technical Design Report - Volume 1: Executive Summary.
ILC-REPORT-2013-040; 2013.
DOI: 10.2172/1347945.

\bibitem{CLIC_CDR}
Linssen L, Miyamoto A, Stanitzki M, et al.
Physics and Detectors at CLIC: CLIC Conceptual Design Report.
CERN Yellow Report CERN-2012-003; 2012.
DOI: 10.5170/CERN-2012-003.

\bibitem{Bai:2021rdg}
Vernieri C, Nanni EA, Dasu S, et al.
A ``Cool'' route to the Higgs boson and beyond: The Cool Copper Collider.
J. Instrum. 2023;18:P07053.
DOI: 10.1088/1748-0221/18/07/P07053.

\bibitem{deFavereau:2013fsa}
DELPHES 3.
DELPHES 3, A modular framework for fast simulation of a generic collider experiment.
J. High Energy Phys. 2014;02:057.

\bibitem{Whizard}
Kilian W, Ohl T, Reuter J, et al.
WHIZARD: Simulating Multi-Particle Processes at LHC and ILC.
Eur. Phys. J. C. 2011;71:1742.

\bibitem{Pythia6}
Sjostrand T, Mrenna S, Skands, et al.
PYTHIA 6.4 Physics and Manual.
J. High Energy Phys. 2006;05:026

\bibitem{Bahr:2008pv}
Bahr M, et al.
Herwig++ Physics and Manual.
Eur. Phys. J. C. 2008;58:639--707.

\bibitem{Bellm:2015jjp}
Bellm J, et al.
Herwig 7.0/Herwig++ 3.0 release note.
Eur. Phys. J. C. 2016;76:196.

\bibitem{Bierlich:2022pfr}
Bierlich C, et al.
A comprehensive guide to the physics and usage of PYTHIA 8.3.
SciPost Phys. Codeb. 2022;2022:8.

\bibitem{Zhu:2022lzv}
Zhu YF, Cui HH, Ruan MQ.
The Higgs\textrightarrow{}$ b\overline{b} $,$ c\overline{c} $, gg measurement at CEPC.
J. High Energy Phys. 2022;11:100.

\bibitem{friedman2001greedy}
Friedman, Jerome H.
Greedy function approximation: A gradient boosting machine.
The Annals of Statistics 2001;29:1189-1232.

\bibitem{chen2017cross}
Chen ZX, Yang Y, Ruan MQ, et al.
Cross section and Higgs mass measurement with Higgsstrahlung at the CEPC.
Chin. Phys. C. 2017;41:023003.

\bibitem{Bai_2020}
Bai Y, Chen CH, Fang YQ, et al.
Measurements of decay branching fractions of $H \to b\bar{b}/c\bar{c}/gg$ in associated $(e^+e^-/\mu^+\mu^-)H$ production at the CEPC.
Chin. Phys. C. 2020;44:013001.

\bibitem{An:2018dwb}
An FF, et al.
Precision Higgs physics at the CEPC.
Chin. Phys. C. 2019;43:043002.

\bibitem{CEPC_Higgs_snowmass2022}
CEPC Physics Study Group.
The Physics potential of the CEPC. Prepared for the US Snowmass Community Planning Exercise (Snowmass 2021).
In: Snowmass 2021, Seattle, WA; 2022.







\end{thebibliography}


\end{document}